\documentclass[review]{elsarticle}

\usepackage{lineno,hyperref,amssymb}
\modulolinenumbers[5]    

\journal{Journal of \LaTeX\ Templates}

\usepackage{amssymb}
\usepackage{amsmath}
\usepackage{latexsym}
\usepackage{epsfig}
\usepackage{graphicx}
\usepackage{color}
\usepackage{fouridx}

\makeatletter

\renewcommand*{\p@subsection}{}

\renewcommand*{\p@subsubsection}{}
\makeatother

\newcommand{\clB}{{\cal B}}
\newcommand{\clR}{{\cal R}}
\newcommand{\clT}{{\cal T}}

\newcommand{\clL}{{\cal L}}
\newcommand{\clH}{{\cal H}}

\newcommand{\hclH}{\hat{\cal H}}

\numberwithin{equation}{section}

\begin{document}

\begin{frontmatter}

\title{Quantum dynamics and relaxation \\ in comb turbulent diffusion \\
{\small\bf   [ Chaos, Solitons \& Fractals 139 (2020) 110305 ]}}

\author{A. Iomin}
\ead{iomin@physics.technion.ac.il}
\address{Department of Physics, Technion, Haifa, 32000, Israel}

\begin{abstract}
Continuous time quantum walks  in the form
of quantum counterparts of turbulent diffusion
in comb geometry are considered.
The interplay between the backbone inhomogeneous
advection $\delta(y)x\partial_x$ along the $x$ axis,
which takes place only at the $y=0$, and normal diffusion
inside fingers $\partial_y^2$ along the $y$ axis
leads to turbulent diffusion.
This geometrical constraint of transport coefficients
due to comb geometry and properties of a dilatation
operator lead to consideration of two possible scenarios
of quantum mechanics. These two variants of continuous time
quantum walks are described by non-Hermitian operators of the form
$\hat{\clH}=\hat{A}+i\hat{B}$.
Operator $\hat{A}$ is responsible for
the unitary transformation, while operator $i\hat{B}$ is responsible
for quantum/classical relaxation. At the first quantum scenario,
the initial wave packet can move against the classical streaming.
This quantum swimming upstream is due to the dilatation operator,
which is responsible for the quantum (not unitary) dynamics
along the backbone,
while the classical relaxation takes place in fingers.
In the second scenario, the dilatation operator is responsible for
the quantum relaxation in the form of an imaginary optical potential,
while the quantum unitary dynamics takes place in fingers.
Rigorous analytical analysis is performed for both
wave and Green's functions.

\end{abstract}

\begin{keyword}
Comb model\sep Dilatation operator\sep
Turbulent diffusion\sep Non-Hermitian Hamiltonian\sep
Log-normal distribution\sep Swimming upstream\sep Imaginary
optical potential
\end{keyword}

\end{frontmatter}

\section{Introduction}\label{sec:int}

We consider a quantum dynamics, which is
a quantum counterpart of turbulent diffusion
in comb geometry.
A possible realization of the turbulent transport
can be discussed
in the framework of a Langevin equation in a
so-called Matheron - de Marsily form \cite{MaMa80},
\begin{equation}\label{MadeMa}
\dot{x_1}=\delta(x_2)vx_1\, \quad \dot{x_2}=\eta(t)\, .
\end{equation}
Here $\eta(t)$ is a random Gaussian delta correlated
process 
$\langle\eta(t)\eta(t')\rangle=2D\delta(t-t')$,
where $D$ is a diffusion coefficient.
In this case, white noise in the $x_2$ axis
affects the velocity of inhomogeneous advection along
the $x_1$ axis at $x_2=0$.
Introducing a probability distribution function (PDF)
$P(x,y,t)= \langle\delta(x_1-x)\delta(x_2-y)\rangle$, we
obtain a Fokker-Planck equation (FPE) in the form of a 2D comb
model \cite{BaIo04}
\begin{equation}\label{comb-eq-1}   
\partial_tP(x,y,t)=-v\delta(y)\partial_x xP(x,y,t)+
D\partial_y^2P(x,y,t)\, .
\end{equation}

The comb Eq. \eqref{comb-eq-1} has been considered in
Ref.~\cite{BaIo04} as a realistic mechanism of
superdiffusion of heavy particles, related to various
applications.
Important role in this turbulent acceleration
of the comb transport and superdiffusion
plays inhomogeneous advection in the form of a
dilatation operator. In the Hamiltonian form,
the dilatation (contraction) operator
\begin{equation}\label{int-01}
\hat{\clH}_0=-ix\hbar\frac{d}{dx}-i\hbar/2\equiv
x\hat{p}_x-i\hbar/2
\end{equation}
attracts much attention
in quantum mechanics and number theory in connection with
the Riemann zeros. This operator has been suggested by Berry
and Keating \cite{BeKe99a,BeKe99b,Connes} in connection to the Riemann
Hypothesis, and $\hclH_0$ is known as a Berry-Keating-Connes
Hamiltonian \cite{Sierra}. In a more general content, the operator has been
considered as a generator of space-time conformal
transformations \cite{AlFuFu76,Jackiw} and entropy problem
in quantum mechanics \cite{vanW}, a random walk approximation to the Riemann
Hypothesis \cite{Armitage}, and the quantum Mellin transform \cite{TwMi06}.
Also it corresponds to quantum dynamics in inverted
potentials \cite{BhKhLa95}, and near hyperbolic points \cite{NoVo97},
including singular behavior \cite{BeVi03} and exponential
spreading of the phase space \cite{Io2013}.
In comb geometry, this operator leads to the exponential
spread of the phase space as well \cite{Io2009}.
Recently, it has been considered as a turbulent diffusion search
optimisation \cite{SaIoKo2020}.

Due to the comb geometry, the backbone inhomogeneous advection
along the $x$ axis acts\footnote{Note that also due to this comb geometry,
inhomogeneous advection is replaced by a random
motion \cite{BaIo04,SaIoKo2019}, see Sec. \ref{sec:Vc}. }
only at the $y=0$.
This geometrical constraint of transport coefficients
leads (at least) to two possible scenarios of quantum mechanical
transport related to turbulent diffusion
described by Eq. \eqref{comb-eq-1}.
In particular,
the comb FPE \eqref{comb-eq-1} has straightforward
relation to quantum mechanics. It is just a one
dimensional quantum mechanical problem
at $y=0$ with the backbone Schr\"odinger equation
$i\hbar\partial_t\psi(x,t)=-iv\hbar x\partial_x \psi(x,t)$.
The classical PDF is mapped on the wave/Green's functions:
$P(x,y=0,t)\rightarrow \psi(x,t)/G_x(x,t)$.
In this case, one can consider
the quantum dynamics governed by the non-Hermitian Hamiltonian
according to the Schr\"odinger equation
\begin{equation}\label{in-sche-1}
i\hbar\partial_t\psi(x,y,t)=\hat{\clH}\psi(x,y,t)
=\left[v\delta(y)\hclH_0 -\tfrac{iD}{\hbar}\hat{p}_y^2\right]\psi(x,y,t)\, .
\end{equation}  
In this case, the quantum (non-unitary) dynamics inside the backbone
along the $x$ axis at $y=0$
is accompanied by relaxation due to diffusion inside fingers
along the $y$ axis.

The second scenario is according to the Wick rotation of time
$t\rightarrow it$ with $D=\hbar/2m$, where $m=1$ is a unit mass of
a quantum particle. Then the non-Hermitian Hamiltonian is
\begin{equation}\label{int-02}
\hclH=\hat{p}_y^2/2+i\hclH_0\delta(y)\equiv \hat{A}+i\hat{B}\, .
\end{equation}
In this case, the backbone is
a quantum trap for the $y$ axis transport, where relaxation is due to
the dilatation operator $\hclH_0$, which works now as a
scattering.
This approach is a generalization of the dynamics of quantum particles
in the presence of traps due to a complex optical potential
\cite{KrLuMa14,DeKhRo2003}.

\section{Turbulent diffusion}\label{sec:Vc}

In this section, we consider the comb Eq. \eqref{comb-eq-1}
as a realistic mechanism of exponential (or turbulent)
superdiffusion and extend a picture of possible
solutions of the problem, described in Ref.~\cite{BaIo04}.
For the completeness of the presentation
we first consider the logarithmic scale of possible solutions.

\subsection{Logarithmic scale}\label{sec:lsv}

Let us consider dynamics for $x(x_1)>0$.
Following to  Ref.~\cite{TwMi06}, we introduce new variable
$w=\ln(x_1)$, and Eq. \eqref{MadeMa}  reads
\begin{equation}\label{lsv-1}
\dot{w}=v\delta(x_2), \quad \dot{x_2}=\eta(t)\, .
\end{equation}
The comb Fokker-Planck equation \eqref{comb-eq-1} in the
log scale for the PDF $P=P(w,y,t)$ now reads
\begin{equation}\label{lsv2}
\partial_tP=-v\delta(y)\partial_wP+D\partial_y^2P
\end{equation}
with the initial condition\footnote{To relate this initial
condition to the initial condition of
Eq. \eqref{comb-eq-1} in the form of Dirac
$\delta$ function, we introduce the multiplier
$1/x_0$ to keep $P_0(x,y)=P_0(w(x),y)=
\frac{1}{x}\delta(\log(x)-\log(x_0))=
\delta(x-x_0)$.}
$P_0(w,y)=\frac{1}{x}\delta(w-w_0)\delta(y)=
e^{-w}\delta(w-w_0)\delta(y)$ and $w\in(-\infty\, ,\infty)$.
After the Laplace transform with respect to $t$, we  look for the
solution in the form
\begin{equation}\label{td-022}
P(w,y,s)=\clL\left[P(w,y,t)\right](s)=f(w,s)g(y,s)=
f(w,s)\exp\left(-|y|\sqrt{s/D}\right)\, .
\end{equation}
This yields
\begin{equation}\label{lsv-3a}
v\partial_wf(w,s)+2\sqrt{Ds}f(w,s)=\frac{1}{x}\delta(w-w_0)\, .
\end{equation} 
We obtain the PDF of the backbone diffusion as follows
\begin{multline}\label{lsv-3}
f(w,t)\equiv P(w,y=0,t)= \\
=\frac{(w-w_0)D}{\sqrt{\pi t^3v^2}}
\exp\left(-\frac{D(w-w_0)^2}{tv^2}\right)e^{-w}\theta(w-w_0)\, .
\end{multline}
In the $x$ space, it corresponds to the
log-normal distribution\footnote{In Eq. \eqref{lsv-4}, we
replace $\theta(w-w_0)$ by $\theta(x-x_0)$ using
a chain of transformations as follows $\frac{d}{dx}\theta(x-x_0)=\delta(x-x_0)=
\frac{1}{x}\delta(w-w_0)=\frac{1}{x}\frac{d}{dw}\theta(w-w_0)
=\frac{d}{dx}\theta(w-w_0)$.}
\begin{equation}\label{lsv-4}
P(x,y=0,t)=\frac{D\ln(x/x_0)}{x\sqrt{\pi t^3v^2}}
\times\exp\left(-\frac{D\ln^2(x/x_0)}{tv^2}\right)\theta(x-x_0)\, .
\end{equation}
Solution in Eq. \eqref{lsv-4} can be extended on $x<0$.
Introducing symmetric initial condition
$\frac{1}{2}[\delta(x-a)+\delta(x+a)]$ at the backbone,
one obtains
\begin{equation}\label{lsv-4a}
P(x,y=0,t)=\frac{\ln(|x|/a)}{2|x|\sqrt{\pi t^3v^2/D}}
\times\exp\left(-\frac{\ln^2(|x|/a)}{tv^2/D}\right)\theta(|x|-a)\, .
\end{equation}

\subsection{Symmetric solution}\label{sec:Ss}

Let us consider the comb equation \eqref{comb-eq-1}
with a symmetrical initial condition at $(x,y)=(\pm a,0)$:
$P_0(x,y)=\delta(y)[\delta(x-a)+\delta(x+a)]/2$, where $a>0$.
Taking into account the Laplace transformation with respect to $t$
and ansatz $P(x,y,s)=f(x,s)\exp\left(-|y|\sqrt{s/D}\right)$,
we obtain
\begin{equation}\label{ss-1}
vx\partial_xf+(2\sqrt{Ds}+v)f=\delta(|x|-a)/2\, ,
\end{equation}
where $x\in (-\infty\, ,\infty)$
and $\delta(|x|-a)\equiv\delta(x-a)+\delta(x+a)$.
The solution of Eq. \eqref{ss-1} is
$f(x,s)=C|x/a|^{-\mu(s)}\theta(|x|-a)$,
where $\mu(s)=2\sqrt{Ds}/v+1$. We also take into account that
$\partial_x|x|=\rm{sgn}(x)=2\theta(x)-1$, and
$\rm{sgn(x)}|x|=x$. An arbitrary constant $C$ is
determined from the normalization condition of $P(x,y,s)$.
Therefore, we obtain a ``dead zone'' in $x\in (-a,a)$ with
$f(x,t)=0$ and log normal distribution according
Eq. \eqref{lsv-4a} for $|x|\ge a$. The ``dead zone'' is due
to the initial condition at $x=\pm a\neq 0$ and
directional outward streaming
from $x=\pm a$ to $x\rightarrow \pm\infty$.
For $a=0$, the solution of Eq. \eqref{ss-1}
reads\footnote{Here we use that $\partial_xx\delta(x)=
\delta(x)+x\delta'(x)=\delta(x)+x(-\delta(x)/x)=0$.}
\cite{BaIo04}
$f(x,s)=C|x|^{1-\mu(s)}+\delta(x)/(\mu(s)-1)$.
In this case, the PDF consists of two parts: the log-normal
distribution, which contributes to the exponentially fast
spreading and the second part is due to the pining delta
function. The latter does not contribute to the mean
squared displacement (MSD).

Correspondingly, the solution for the PDF $P(x,y,t)$ is a
convolution integral
\begin{equation}\label{ss-2}
P(x,y,t)=P(x,y=0,t)\ast P_y(y,t)
=\int_0^tP(x,y=0,t-t')P_y(y,t')dt'\, ,
\end{equation}
where $P(x,y=0,t)=f(x,t)$ and $P_y(y,t)$ is
the L\'evy-Smirnov density defined by the inverse Laplace
transform,
\begin{equation}\label{ss-3}
P_y(y,t)=\clL^{-1}\left\{\exp[-|y|\sqrt{s/D}]\right\}\, .
\end{equation}

\section{Quantum dynamics: Eigenvalue's expansion}\label{sec:Ee}

We consider two scenarios of the quantum dynamics
described by non-Hermitian operators.
The first one is the quantum dynamics along the backbone
with relaxation in the form of classical diffusion along the fingers.
Contrary to that, the second scenario is the quantum dynamics in fingers
while relaxation in the backbone is due to the dilatation operator.
In both cases, the dilatation operator $H_0$ plays
important role for the quantum tasks.
Therefore we start from the eigenvalue problem of the dilatation operator.

The dilatation operator \eqref{int-01}
$$\hat{\clH}_0=\hbar [-ix\partial_x-i/2]=\hat{p}_xx+i\hbar /2$$
determines the complete set of eigenfunctions $\chi_{\omega}(x)$ with
the eigenvalues $\omega$ according to the eigenvalue problem
$\hat{\clH}\chi_{\omega}(x)=\hbar\omega\chi_{\omega}(x)$,
where $\omega$ is the continuous spectrum and the
eigenfunctions are \cite{BeKe99a,Sierra}
\begin{equation}\label{Ee1}   
\chi_{\omega}(x)=\frac{1}{\sqrt{N|x|}}
\exp[i\omega\ln|x|] \,,
\end{equation}
which satisfies the boundary conditions
$\chi_{\omega}(x=\pm\infty)=0$ and $ N=4\pi$.
For the continuous spectrum, the normalization
condition is
\begin{equation}\label{Ee2}   
\int_{-\infty}^{\infty}\chi_{\omega'}^*(x)\chi_{\omega}(x)dx
=\delta(\omega-\omega')\, ,
\end{equation}
while the completeness relation is
$\int \chi_{\omega}^*(x')\chi_{\omega}(x)d\omega=\delta(x-x')$
(see {\em e.g.} \cite{LL}). Note also that for $x>0$ the
normalization constant is half as larger, $N=2\pi$
\cite{Sierra}.

It is worth mentioning that a mathematically rigorous calculation of the
normalization constant for the wave function $\chi_{\omega}(x)$
can be presented by following\footnote{We present these arguments
from Ref. \cite{Io2013} with necessary corrections.} the presentation due to
the monograph by V.A. Fock \cite{fok}. Since the operator $ x\hat{p}$ has
continuous spectrum  $\omega$, the eigenfunctions
$\chi(\omega,x)\equiv\chi_{\omega}(x)$ are not square
integrable. Therefore, the normalization condition exists not for
the eigenfunction but for the ``eigendifferential'' \cite{fok}
$\Delta\chi(\omega,x)$, which reads
$ \Delta\chi(\omega,x)=\int_{\omega}^{\omega+\Delta\omega}
\chi(\omega',x)d\omega'$.
Substituting here Eq. \eqref{Ee1}, one obtains
$$ \Delta\chi(\omega,x)=\frac{2}{\sqrt{N|x|}\ln|x|}
\exp[i(\omega+\Delta\omega)\ln|x|]\sin\frac{\Delta\omega\ln|x|}{2}\, .$$
This solution is already square integrable and has the
normalization form
\begin{equation}\label{Ee3}
\lim_{\Delta\omega\to 0}\frac{1}{\Delta\omega}
\int_{-\infty}^{\infty}dx\left|\Delta\chi(\omega,x)\right|^2=1\, .
\end{equation}
To take the limit in Eq. \eqref{Ee3}, the integrand can be presented as follows
$$\left|\Delta\chi(\omega,x)\right|^2=\frac{4}{N|x|\ln^2|x|}
\sin\frac{\Delta\omega\ln|x|}{2}\sin\frac{\Delta\Omega\ln|x|}{2}\, ,$$
where $\Delta\Omega=\Delta\omega+\Delta\omega_1$.
We can perform this trick, since
the ``eigendifferentials'' for not overlapping spectral regions $\Delta\omega$ and $\Delta\omega_1$ are orthogonal. \cite{fok}.
Then taking the limit
$\Delta\omega=0$ and carrying out the variable change
$z=(\Delta\omega_1/2)\ln |x|$ and taking into account that
$\frac{4}{N}\int_{-\infty}^{\infty}\frac{\sin z}{z}dz=\frac{4\pi}{N}$,
we obtain $N=4\pi$, which coincides exactly with the dimensionless
normalization constant $N$ in Eq. \eqref{Ee1}.

\section{Quantum dynamics in backbone}\label{sec:bQD}

The Schr\"odinger equation \eqref{in-sche-1} for the Green's
function, described the dynamics of the first scenario, is
\begin{equation}\label{bQD-1}
i\hbar\partial_tG(x,y,t)
=v\delta(y)(x\hat{p}_x-i\hbar/2)G(x,y,t)
-\tfrac{iD}{\hbar}\hat{p}_y^2G(x,y,t)
\end{equation}
with the initial condition $G(x,y,t=0)=\delta(x-x_0)\delta(y)$
and $D=\hbar/2$.
In the Laplace domain $\clL[\psi(t)](s)=\tilde{\psi}(s)$,
Eq. \eqref{bQD-1} reads
\begin{multline}\label{bQD-2}
is\hbar \tilde{G}(x,y,s)=
v\delta(y)(x\hat{p}_x-i\hbar/2)\tilde{G}(x,y,s) \\
-\tfrac{iD}{\hbar}\hat{p}_y^2\tilde{G}(x,y,s)
-i\hbar\delta(x-x_0)\delta(y)\, .
\end{multline}
The Laplace image of the Green's function
has the multiplicative form of both backbone
and finger's dynamics
\begin{equation}\label{bQD-3}
\tilde{G}(x,y,s)=\tilde{G}(x,y=0,s)g(y,s)=
f(x,s)e^{-|y|\sqrt{s/D}}\, ,
\end{equation}
where $g(y,s)=e^{-\sqrt{s/D}|y|}$ describes relaxation of
quantum dynamics in the form of classical
diffusion in fingers, while $f(x,s)$ describes
the quantum dynamics in the backbone.
Substituting solution \eqref{bQD-3} in Eq. \eqref{bQD-2},
we obtain equation for the backbone quantum dynamics
\begin{equation}\label{bQD-4}
-iv(x\partial_x+\tfrac{1}{2})f(x,s)-i\sqrt{2\hbar s}f+i\delta(x-x_0)=0.
\end{equation}

Formally, it corresponds to the PDF of classical diffusion,
described in Sec. \ref{sec:Ss}, with
$\mu(s)=1+\sqrt{2\hbar s}/v$. When $x_0=0$, the solution reads
\begin{equation}\label{bQD-5}
f(x,s)=|x|^{\tfrac{1}{2}-\mu(s)} +\delta(x)/[\mu(s)-1/2]\, .
\end{equation}

Further quantum mechanical analysis is carried out
in the framework of the spectral decomposition according to
the eigenvalue problem of Sec. \ref{sec:Ee}.
Since $\chi_{\omega}(x)$ is a complete set
of eigenfunction, $f(x,s)$ can be considered as a superposition
\begin{equation}\label{bQD-6}
f(x,s)=\int d\omega\chi_{\omega}(x)b_{\omega}(s)\, .
\end{equation}
From Eq. \eqref{bQD-4}, this yields equation for
$b_{\omega}(s)\equiv b(s) $ as follows
\begin{equation}\label{bQD-7}
v\omega b(s)-i\sqrt{2\hbar s}b(s) =-i\chi^*_{\omega}(x_0)\, .
\end{equation}
Resolving this equation with respect to $b(s)$ and substituting it in Eq.
\eqref{bQD-6}, we obtain
\begin{multline}\label{bQD-8}
f(x,s)=\frac{1}{iv}\int\frac{ d\omega\chi_{\omega}(x)\chi^*_{\omega}(x_0)}
{\omega-i\sqrt{2\hbar s}/v} =\\
=\frac{1}{4\pi iv|x|}\int\frac{d\omega e^{i\omega\ln|x/x_0|}}
{\omega-i\sqrt{2\hbar s}/v}= \\
=\frac{1}{2v|x|}e^{-\sqrt{2\hbar s}\ln|x/x_0|/v} \, ,
\end{multline}
where $x>x_0$ for $\rm{Re}\sqrt{s} >0$ and $x<x_0$ if $\rm{Re}\sqrt{s} <0$.
We note that there is no anymore the ``dead zone'', and
this quantum situation changes drastically from classical
turbulent diffusion discussed in Sec. \ref{sec:Vc}.
We call this effect a
\textit{quantum swimming upstream}\footnote{Note that
here we have only an indication on this phenomenon.
A more accurate and elegant way to obtain this result
will be carried out in Sec. \ref{sec:FF}.}.

Substituting the result of Eq. \eqref{bQD-8} in Eq. \eqref{bQD-3},
we obtain the Green's function in the form of the Laplace inversion
\begin{multline}\label{bQD-9}
G(x,y,t)=\frac{1}{2v|x|}
\clL^{-1}\left[e^{-(v|y|+\ln|x/x_0|)\sqrt{2s\hbar}/v}\right]= \\
=\frac{(v|y|+\ln|x/x_0|)}{4v^2|x|\sqrt{2\pi t^3/\hbar}}
\exp\left[-\hbar\frac{(v|y|+\ln|x/x_0|)^2}{2v^2t}\right]\, ,
\end{multline}
which is the quantum Green's function in the form of the
combination of the log-normal and L\'evy-Smirnov
densities.

\subsection{Fractional dynamics in backbone}\label{sec:Dis1}

Let us show that behind this classical
L\'evy-Smirnov form of the Green's function
there is a fractional quantum dynamics along the backbone.
To this end, let us perform the Laplace inversion
before performing integration with respect to the
spectrum in Eq. \eqref{bQD-8}.
Then the expansion coefficient
$b_{\omega}(s)$ in Eq. \eqref{bQD-7} reads
\begin{equation}\label{Dis1-1}
 b_{\omega}(s) = \frac{A_{\omega}(x_0)}{\sqrt{s}+i\tilde{\omega}}\, ,
\end{equation}
where $\tilde{\omega}=v\omega/\sqrt{2\hbar}$ and
$A_{\omega}(x_0)=\chi^*_{\omega}(x_0)/\sqrt{2\hbar}$.
Performing the Laplace inversion, we obtain the
temporal behavior of the coefficients in the form of the
Mittag-Leffler functions $E_{\alpha,\beta}(z)$ \cite{BaEr55}:
\begin{equation}\label{Dis1-2}
B_{\omega}(t)=\clL^{-1}[b_{\omega}(s)](t)=A_{\omega}(x_0)
t^{-\frac{1}{2}}
E_{\frac{1}{2},\frac{1}{2}}\left(-i\tilde{\omega}t^{\frac{1}{2}}\right)\, .
\end{equation}
Therefore, the backbone component of the Green's function reads
\begin{multline}\label{Dis1-3}
f(x,t)=\int B_{\omega}(t)\chi_{\omega}(x)d\omega = \\
=\frac{1}{\sqrt{2\hbar t}}\int d\omega \chi^*_{\omega}(x_0)
E_{\frac{1}{2},\frac{1}{2}}\left(-i\tilde{\omega}t^{\frac{1}{2}}\right)
\chi_{\omega}(x) = \\
=\frac{1}{\sqrt{2\hbar t}}
E_{\frac{1}{2},\frac{1}{2}}\left(-i\hat{H}t^{\frac{1}{2}}\right)
\delta(x-x_0)\equiv G_{x}\left(i\hat{\clH}t^{\frac{1}{2}}\right)\, ,
\end{multline}
where the ``Hamiltonian''
is $\hat{\clH}= \tfrac{v\hclH_0}{\sqrt{2\hbar}}$.

The Green's function is the convolution of two non-unitary
dynamics, which are the quantum fractional dynamics along the backbone
with the Green's function $G_x(x,t)=f(x,t)$
and the finger's dynamics according to the L\'evy-Smirnov density
\begin{equation}\label{Levy-Smirnov}
G_y(y,t)=
\clL^{-1}\left[\exp\left(-|y|\sqrt{2s\hbar}\right)\right](t)
=\frac{\hbar|y|}{\sqrt{\pi t^3}}
\exp\left(-\hbar\frac{y^2}{2t}\right)\, .
\end{equation}
That is
\begin{equation}\label{Dis1-4}
G(x,y,t)=G_x\left(\hat{\clH}t^{\frac{1}{2}}\right)\ast G_y(y,t),
\end{equation}
which yields the result in Eq. \eqref{bQD-9}.
Therefore, the quantum dynamics at the backbone is not unitary.

\section{Fox H function for quantum swimming upstream}\label{sec:FF}

The Mittag-Leffler function can be presented in the form of the Fox
$H$ function  \cite{MaHa08} (see also Exampl. 2.7
in Ref. \cite{IMH2018}) defined in terms of the Mellin-Barnes integral,
\begin{equation}\label{FF-1}
E_{\alpha,\beta}(-z)=H^{1,1}_{2,1}(z)=\frac{1}{2\pi i}
\int_C\frac{\Gamma(\xi)\Gamma(1-\xi)}{\Gamma(\beta-\alpha\xi)}
(z)^{-\xi}d\xi\, ,
\end{equation}
where $\Gamma(\xi)$ is a gamma function and the contour $C$ starts
at $c-i\infty$, ends at $c+i\infty$ and separates the
poles of the gamma functions $\Gamma(\xi)$ and $\Gamma(1-\xi)$, such that
$0<\rm{Re}(\xi)<1$. In Eq. \eqref{FF-1},
$z=i\tilde{\omega}t^{\frac{1}{2}}$ is the imaginary variable
and $\alpha=\beta=1/2$.
Integration with respect to the spectrum reads
\begin{equation}\label{FF-2}
I(x/x_0)=\frac{1}{4\pi |x|}\int_{-\infty}^{\infty}
(i\omega)^{-\xi} e^{i\omega\ln|x/x_0|} d\omega\, .
\end{equation}
To perform this integration, the variable change
$i\omega=\tau$ yields the standard Laplace inversion
\begin{equation}\label{FF-3}
I(x/x_0)=\frac{1}{2|x|}\cdot\frac{1}{2\pi i}
\int_{-i\infty}^{i\infty}\tau^{-\xi}e^{w\tau}d\tau
=\frac{w^{\xi-1}}{2|x|\Gamma(\xi)}\, ,
\end{equation}
where $w\equiv w(x/x_0)=\ln|x/x_0|>0$ and $0<\rm{Re}(\xi)<1$.
When $w<0$, it means that a quantum particle is in
the dead zone $|x|<|x_0|$, then the variable change is
$\tau=-i\omega$. The integral in Eq. \eqref{FF-3} reads
\begin{equation}\label{FF-4}
I(x/x_0)=\frac{1}{2|x|}\frac{1}{2\pi i}\int_{-i\infty}^{i\infty}
(-\tau)^{-\xi} e^{\tau|w|} d\tau
=\frac{e^{-i\pi\xi}|w|^{\xi-1}}{2|x|\Gamma(\xi)}
=-\frac{w^{\xi-1}}{2|x|\Gamma(\xi)}\, .
\end{equation}
Therefore, taking into account results in Eqs.
\eqref{FF-1} - \eqref{FF-4} for the integration \eqref{Dis1-3}.
we obtain the backbone Green's function as follows
\begin{multline}\label{FF-5}
G_x(x,t)\equiv f(x,t)=
\int B_{\omega}(t)\chi_{\omega}(x)d\omega = \\
=\pm
\frac{w^{-1}}{2|x|\sqrt{2\hbar t}}
\frac{1}{2\pi i} \int_C \frac{\Gamma(1-\xi)}
{\Gamma(\frac{1}{2}-\frac{1}{2}\xi)}
\left(\frac{vt^{\frac{1}{2}}}{w\sqrt{2\hbar}}
\right)^{-\xi}d\xi \, ,
\end{multline}
where $(+)$ sign is for $w>0~(|x/x_0|>1)$ and
$(-)$ sign is for $w<0~(|x/x_0|<1)$.
Now, using the Legendre duplication formula,
$\Gamma(z)\Gamma(z+1/2)=2^{1-2z}\sqrt{\pi}\Gamma(2z)$,
which yields $\tfrac{\Gamma(1-\xi)}{\Gamma(1/2-\xi/2)}
=2^{-\xi}\Gamma(1-\xi/2)/\sqrt{\pi}$.
Defining $p=1-\xi/2$, we obtain Eq. \eqref{FF-5}
as follows (see \textit{e.g.}, Eq. (2.43) in Ref. \cite{IMH2018})
\begin{multline}\label{FF-6}
G_x(x,t)=
\pm \frac{w^{-1}}{2|x|\sqrt{2\pi\hbar t}} \\
\times \frac{\hbar w^2}{2v^2 t}
\frac{1}{2\pi i} \int_C\left(\frac{\hbar w^2}{2v^2t}\right)^{-p}
\Gamma(p)dp = \\
=\pm \frac{\ln\left|\tfrac{x}{x_0}\right|}{4|x|v^2\sqrt{2\pi t^3 /\hbar}}
\exp\left[-\hbar\frac{\ln^2\left|\tfrac{x}{x_0}\right|}{2v^2t}\right]\, .
\end{multline}
The Green's function in Eq. \eqref{FF-6} is a more accurate result on the quantum dynamics inside the ``dead zone''  for $ |x|<|x_0|$. It
contains the phase multiplier $e^{-i\pi}$, which corresponds to the
quantum swimming upstream.

\section{Quantum dynamics in fingers}\label{sec:QDF}

In this section we consider the second scenario,
which is performed according to the Wick rotation of
time $t\rightarrow it$ in Eq. \eqref{comb-eq-1}
with $D=\hbar/2m$, and $m=1$ is a unit mass of
a quantum particle. Then the Schr\"odinger equation
reads
\begin{equation}\label{QDF-1}
i\hbar\partial_t\psi(x,y,t)=\hclH\psi(x,y,t)\, ,
\end{equation}
where the non-Hermitian Hamiltonian is
$\hclH=\hat{p}_y^2/2+iv\hclH_0\delta(y)$.
An important point of the analysis should be
admitted: it is the imaginary optical potential,
which acts as a trap \cite{KrLuMa14,DeKhRo2003,AgMuBl10}.
In the finger's quantum dynamics the trap relates
to the dilatation operator and leads to the imaginary delta
function, see also \ref{sec:App-A}.
To understand this phenomenon, the dynamics of both the wave and Green's functions is analysed.

Again, we consider the wave function as a decomposition over
the complete set $\chi_{\omega}(x)$,
\begin{equation}\label{QDF-2}
\psi(x,y,t)=\int \phi_{\omega}(y,t)\chi_{\omega}(x)d\omega\, .
\end{equation}
Then taking into account properties of the dilatation/contraction
operator $\hclH_0$ and owing to Eq. \eqref{QDF-2}, we present the initial
condition as the spectral decomposition as well with the
Gaussian weight
$\rho_a(\omega)=\left[\frac{2a}{\pi}\right]^{\frac{1}{4}}\exp(-a\omega^2)$
with real $a>0$  \cite{Io2013}.
This yields the initial condition for the backbone
wave function in the form of a log-normal distribution.
Eventually, we obtain the initial condition for $\phi_{\omega}(y,t=0)$
as follows
\begin{equation}\label{QDF-2a}
\phi_{\omega}(y,t=0)=\rho_a(\omega)\Phi_0(y,\omega)\, ,
\end{equation}
where the choice of $\Phi_0(y,\omega)$ depends on the choice of
the boundary conditions for every $\omega$ mode.
In particular, free boundary conditions is considered in what follows.

\subsection{Wave function}\label{sec: FreeBC}

In the case of free conditions at the boundaries of the
fingers, we arrive at the $y$ transport in the form
of a scattering problem with the imaginary potential
$iv\hbar\omega\delta(y)$ as follows
\begin{equation}\label{QDF-3}
i\hbar\partial_t\phi_{\omega}=-\frac{\hbar^2}{2}\partial_y^2\phi_{\omega}
+iv\hbar\omega\delta(y)\phi_{\omega}\, .
\end{equation}
Looking for the eigenvalue solutions
$\phi_{\omega}(y,t)=e^{-iet/\hbar}\Phi_e(y)$,
we arrive at the eigenvalue problem
\begin{equation}\label{QDF-4}
-\frac{\hbar^2}{2}\partial_y^2\Phi_e
+i\lambda\delta(y)\Phi_e=e\Phi_e\, ,
\end{equation}
with free boundary conditions at infinities
and $\lambda=\hbar v\omega$.
Contrary to the real (attractive) delta potential \cite{Bl88},
there are no bound states. 
However, still we can consider two classes of the
spectrum \cite{IMH2018,Bl88}.
The first class corresponds to the scattering problem and contains the
continuous spectrum solutions, which are standing waves. The solutions
can be considered as the result of
the scattering by the $i\delta$ potential of the left and right
incident waves \cite{Bl88},
\begin{equation}\label{QDF-5}  
\Phi^{\pm}_k(y)=\frac{1}{\sqrt{2\pi}}\left[e^{\pm iky}+
\mathcal{B}(k)e^{ik\lvert y\rvert}\right],
\end{equation}
where $\mathcal{B}(k)={\lambda}/{(\hbar^2k-\lambda)}$ is the
``scattering amplitude'', determined by Eq.\ \eqref{QDF-4}
and related to imaginary potential\footnote{In contrast to
the scattering amplitude of a real potential, here the scattering amplitude
does not correspond to the optical theorem,
see \ref{sec:App-A}.},  and $e=e(k)=\hbar^2k^2/2$.

In further analysis we are interesting in the second
class of the solutions, which is but one (not bound) state
\begin{equation}\label{QDF-6}
\Phi_0(y)=(-i\lambda)^{1/2}e^{i\lambda|y|}\, , \quad e_0=\lambda^2/2\, .
\end{equation}
We choose $\Phi_0(y,\omega)=\Phi_0(y)$  to be the initial condition
for the finger's quantum dynamics.
This immediately yields the
quantum evolution of the finger's wave function
\begin{equation}\label{QDF-7}
\phi_{\omega}(y,t)=(-i\hbar\omega v)^{\frac{1}{2}}
e^{-i\frac{\hbar}{2}(\omega v)^2t}
\cdot e^{i\hbar\omega v|y|}\,
\end{equation}
and the dynamics (including the initial condition)
of the comb wave function reads from Eqs. \eqref{QDF-2} and \eqref{QDF-2a},
and \eqref{QDF-7} as follows
\begin{equation}\label{QDF-8}
\psi(x,y,t)
=\frac{(-i\hbar v)^{\frac{1}{2}}}{\sqrt{4\pi |x|}}
\left[\frac{2a}{\pi}\right]^{\frac{1}{4}}
\times\int_{-\infty}^{\infty}
\frac{e^{-A(t)\omega^2}}{\omega^{\frac{1}{2}}}
e^{i\omega (\hbar v|y|+\ln|x|)}d\omega\, ,
\end{equation}  
where $A(t)=a+i\hbar v^2t$. When $t=0$ it corresponds to the
initial condition of the comb wave function.
The exact solution of the wave function is expressed in the form
of the hypergeometric and Bessel functions.
However it is instructive to present the result in the form of
elementary functions\footnote{We consider
the table integral \cite{BaEr54} in Eq. \eqref{QDF-8} as follows
$\int z^{-\frac{1}{2}}e^{-Az^2}e^{iBz}dz=
(1-i)2^{-1}A^{\frac{1}{4}}\Gamma(\tfrac{1}{4}){}_1F_1[\frac{1}{4};
\frac{1}{2};-\frac{B^2}{4A}]+(i-1)2^{\frac{3}{2}}\pi B^{\frac{1}{2}}
e^{-\frac{B^2}{8A}} I_{\frac{1}{4}}\left(\frac{B^2}{8A}\right)$.
The stationary phase approximation for the integral yields
$\left[\frac{2\pi}{iB}\right]^{\frac{1}{2}}e^{-\frac{B^2}{2A}}$.}.
To this end, the integral in Eq. \eqref{QDF-8} is evaluated by the
stationary phase approximation, which yields
\begin{equation}\label{QDF-9}
\psi(x,y,t)
=\frac{i(\hbar v)^{\frac{1}{2}}}{\sqrt{|x|(\hbar v|y|+\ln|x|)}}
\cdot\left[\frac{a}{2\pi}\right]^{\frac{1}{4}}
\times\exp\left[-\frac{(\hbar v|y|+\ln|x|)^2}{(a+i\hbar v^2 t)}\right]\, .
\end{equation}

\subsection{Green's function}\label{sec:Dis2}

\begin{figure}[htbp]
\includegraphics[width=1.0\hsize]{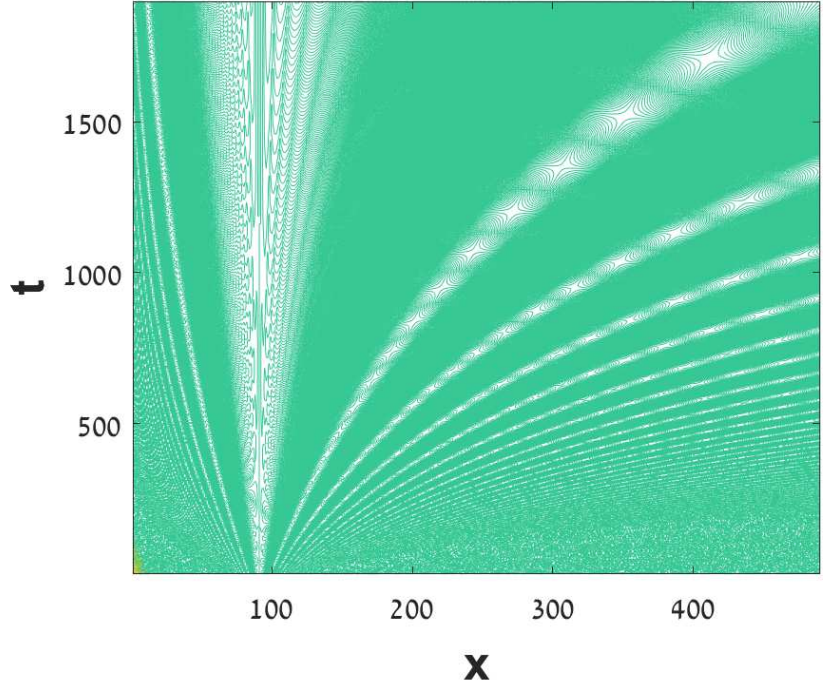}
\caption{Contour presentation of the real part of
Green's function \eqref{Dis2-3}
as a function of $t$ and $x$ for $t\in[10^{-5}\, ,  2\cdot 10^{-4}]$ and
$x\in [10^{-1}\cdot x_0\, ,5\cdot x_0] $, where the initial condition $x_0=10^{-3}$.
The imaginary part exhibits the same $x-t$ structure.
All parameter and constants are taken in dimensionless form
such that $\sqrt{\hbar}=2v$. The contour
is plotted on $x-t$ grid with $\Delta x=10^{-5}$ and $\Delta t=10^{-7}$.
In the case the axes are measured the number of steps on the grid.
At the initial stage, for small
$x$ and $t$ its behavior is random oscillations. Then it reflexes a
complicated quasi-periodic structure of the oscillations.
The white color corresponds to the positive values of $\mathrm{Re}[G_x]$,
while the turquoise color corresponds to the negative values.
However, the turquoise color regions are more colorful as zoom shows in
Fig.~\ref{fig:fig2}. }
\label{fig:fig1}
\end{figure}

\begin{figure}[htbp]
\includegraphics[width=1.0\hsize]{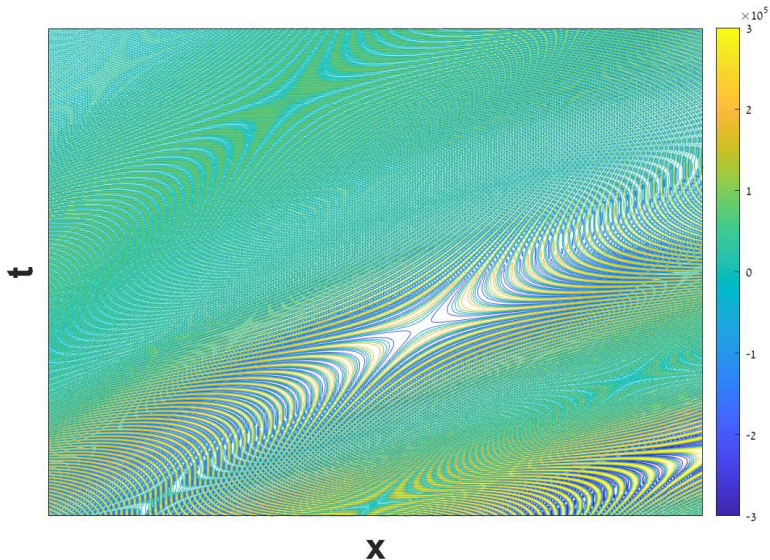}
\caption{Zoom of a part of the region in  Fig.~\ref{fig:fig1},
where $x/x_0\in[1.5\, , 2]$ and $t\in [4\cdot 10^{-5}\, , 7\cdot 10^{-5}]$.
More details are shown with smaller cells of the grid:
$\Delta x=2\cdot 10^{-6}$ and $\Delta t=5\cdot 10^{-8}$}
\label{fig:fig2}
\end{figure}

Let us analyse the Green's function $G(x,y,t)$,
in particular the Green's function
of the backbone $G_x(x,t)$ in spirit of Sec. \ref{sec:bQD}.
Rewriting Eq. \eqref{QDF-1} for the Laplace image of the
Green's function  $\clL[G(x,y,t)](s)=\tilde{G}(x,y,s)=
\tilde{G}_x(x,s)\tilde{G}_y(y,s)$, we have
\begin{equation}\label{Dis2-1}
i\hbar s \tilde{G}=-\frac{\hbar^2}{2}\partial_y^2\tilde{G}
+iv\hclH_0\delta(y)\tilde{G} +i\hbar\delta(x-x_0)\delta(y)\, .
\end{equation}
Now, the ansatz for the finger Green's
function is $\tilde{G}_y(y,s)=e^{i\sqrt{2si/\hbar}}$,
while the backbone Green's function is a superposition
\eqref{bQD-6}, namely
$\tilde{G}_x(x,s)=\int b_{\omega}(s)\chi_{\omega}(x)$.
Taking these expressions into account in Eq. \eqref{Dis2-1},
we obtain the expression for the Green's function $G_x(x,t)$
in the form of Eq. \eqref{bQD-8} as follows
\begin{multline}\label{Dis2-2}
\tilde{G}_x(x,s)=\frac{-1}{v}\int
\frac{d\omega \chi_{\omega}(x) \chi^*_{\omega}(x_0)}
{\omega-i(1-i)\sqrt{\hbar s}/v} =\\
=\frac{-1}{4\pi v|x|}\int\frac{d\omega e^{i\omega\ln|x/x_0|}}
{\omega-i(1-i)\sqrt{\hbar s}/v}= \\
=\frac{-i}{2v|x|}e^{-(1-i)\sqrt{\hbar s}\ln|x/x_0|/v} \, .
\end{multline}
Performing the Laplace inversion we obtain
\begin{equation}\label{Dis2-3}
G_x(x,t)=\frac{\sqrt{i\hbar}\ln\left|\frac{x}{x_0}\right|}
{2v|x|\sqrt{2\pi t^3}}
\exp\left[\frac{i\hbar\ln^2\left|\frac{x}{x_0}\right|}{2v^2t}\right]\, .
\end{equation}
The behaviour of this solution is extremely complicated: it locks
completely random at the initial time and space and then changes to
complicated quasi-periodic behavior. The contour plot of $G_x(x,t)$
versus $(x,t)$ is shown in Figs. \ref{fig:fig1} and \ref{fig:fig2},
where Fig. \ref{fig:fig2} is a zoom of part of the $(x,t)$ space.

\section{Conclusion}

Two examples of quantum dynamics is considered in
the form of quantum counterparts
of turbulent diffusion in comb geometry. The latter is described by
either Langevin equation in the Matheron - de Marsily
form of Eq. \eqref{MadeMa} or equivalently by the Fokker-Planck equation
(FPE) in the form of the 2D comb model \eqref{comb-eq-1}.
The solution of the latter form corresponds to the
log-normal-L\'evy-Smirnov density
of the random variable $x$ with the variance growing exponentially:
$\mathrm{var}(x)=\langle (x-\langle x\rangle)^2\rangle\sim e^t$,
It describes relative diffusion of a pair of particles,
namely the distance between them. Therefore one finds
that the relative diffusivity of any two particles grows
with the inter-particle distance; and due to this growth,
this process has a property of turbulent diffusion.

Having concerned with the quantum dynamics, we have constructed
two variants of continuous time quantum walks related
to non-Hermitian operators of the form
$\hat{\clH}=\hat{A}+i\hat{B}$, where $\hat{A}$ is responsible for
the unitary transformation, while $i\hat{B}$ is responsible
for quantum/classical relaxation.
The first model in Eq. \eqref{bQD-1} is obtained by multiplication
of the comb model by $\hbar\sqrt{-1}$. As a rule, such simple
multiplication of an equation by a constant does not lead to a
new solution. However, due to the forms of the classical advection
$x\partial_x$ and the dilatation operator $\hat{\clH}_0=
-i\hbar(x\partial_x+1/2)=\hat{A}$, one immediately arrives at
non-Hamiltonian quantum mechanics inside the backbone,
which leads to quantum effects. In particular,
it is quantum tunneling inside a dead zone, which is a
quantum swimming upstream, that is a quantum initial
wave packet moves against
the classical streaming. The fingers play a role of classical environment
with classical relaxation according to normal diffusion.

The second model corresponds to a standard quantum counterpart
of a classical FPE by the Wick rotation of time
$t\rightarrow it$. The unitary dynamics inside fingers is due to
free motion, $\hat{A}=\frac{\hbar^2}{2}\partial_y^2$. The
relaxation is due to the dilatation operator $\hat{B}=
\hat{\clH}_0\delta(y)$, which is an imaginary optic
scattering potential for the incident plain wave coming from the fingers.
It is also a trap, which violets the optical theorem.

In conclusion, we comment symmetrical-asymmetrical positioning of the initial conditions, which is a generic task related to the comb geometry for both quantum mechanics and classical diffusion.
To be specific, let us consider the initial condition for Eq. \eqref{comb-eq-1} in a finger with $y=y_0\neq 0$  with the initial PDF $P_0(x,y)=\delta(x-x_0)\delta(y-y_0)$.
Then, following Eq. \eqref{td-022} with $g(y,s)=e^{-|y|\sqrt{s/D}}$, the finger transport is started first
with the L\'evy-Smirnov density \eqref{ss-3},
$P_y(y-y_0,t)=\tfrac{|y-y_0|}{\sqrt{4D\pi t^3}}
e^{-\frac{(y-y_0)^2}{4Dt}}$,
which is also
a first arrival time distribution to the backbone \cite{SaIoKo2019}.
Then, in the backbone's transport equation,
the L\'evy-Smirnov distribution $P_y(y_0, t)$ plays a role of a
strength of a $\delta$ source - sink term at $x = x_0$.
This situation differs from both scenarios considered above
and has been discussed for anomalous diffusion and random
search in two and three dimensional combs \cite{IoSa_MDPI,LSRJIK2020}.

\appendix

\section{Imaginary optical potential vs optical theorem}\label{sec:App-A}

\def\theequation{A.\arabic{equation}}
\setcounter{equation}{0}

The optical theorem, results from the conservation
of the probability flux at elastic scattering
and establish the relation between a scattering cross section and
a scattering amplitude, and it also relates to the interference between
incident and scattered waves \cite{PeZe98}.
Some general aspects of the one dimensional scattering,
including the optical theorem, have been considered as well
\cite{Eb65}.
To understand the violation of the optical theorem
by scattering at imaginary optical potential
$i\delta(y)$, we first consider Eq. \eqref{QDF-4}
for the real attractive delta potential \cite{IMH2018,Bl88},
which reads
\begin{equation}\label{A1}
-\frac{\hbar^2}{2}\partial_y^2\Phi_e
-\lambda\delta(y)\Phi_e=e\Phi_e\, .
\end{equation}
The scattering solutions \eqref{QDF-5}
for the left and right incident waves are
\begin{equation}\label{A-2}  
\Phi^{\pm}_k(y)=\frac{1}{\sqrt{2\pi}}\left[e^{\pm iky}+
\mathcal{B}(k)e^{ik\lvert y\rvert}\right],
\end{equation}
where the scattering amplitude and the energy are
\begin{equation}\label{A-3}
\mathcal{B}(k)=\frac{-\lambda}{(\lambda-i\hbar^2k)}\, ,
\quad \quad e=e(k)=\hbar^2k^2/2\, .
\end{equation}
Let us consider the flux, specifically for the left incident
wave $\psi_k(y)=\Phi^{+}_k(y)$ it reads
\begin{multline}\label{A-4}
\frac{J_k(y)}{\hbar k}=\frac{1}{2i}\left[\psi_k^{*}\partial_y\psi_k
-\psi_k\partial_y\psi_k^{*}\right]
=1-
\frac{\lambda(1+\rm{sgn}(y))e^{ik(|y|-y)}}{2(\lambda-i\hbar^2k)}
\\
-\frac{\lambda(1+\rm{sgn}(y))e^{-ik(|y|-y)}}{2(\lambda+i\hbar^2k)}
+\frac{\lambda^2\rm{sgn}(y)}{|\lambda+i\hbar^2k|^2}] =\\
=1+(1+\rm{sgn}(y))\rm{Re}\mathcal{B}(k)\cos[k(|y|-y)]
+\left|\mathcal{B}(k)\right|^2\rm{sgn}(y)\, ,
\end{multline}
where $(1+\rm{sgn}(y))\sin[k(|y|-y)]\equiv 0$
and $\rm{Re}\mathcal{B}(k)=-\left|\mathcal{B}(k)\right|^2$.
The letter expression is nothing more as an analog of
the optical theorem relevant for the one dimensional
scattering \cite{Eb65}.

Note that the scattering amplitude $\mathcal{B}(k)$
in the solution \eqref{A-2} determines the reflection and
transmission coefficients. Namely, for $y<0$, the reflection
coefficient is $\clR=\mathcal{B}(k)$, while
the transmission coefficient for $y>0$ is
$\clT=1-\clB$. This immediately yields
the law of conservation of the number of particles,
$|\clT|^2+|\clR|^2=1$.
From Eq. \eqref{A-4}, also follows that
$J_k(y>0)=J_k(y<0)=\hbar k-
\hbar k\left|\mathcal{B}(k)\right|^2 $.

Now let us consider the current for the scattering at
the imaginary optical potential $i\delta(y)$.
From the solution \eqref{QDF-5} with
$\mathcal{B}(k)={\lambda}/{(\hbar^2k-\lambda)}$, we have
that now the scattering amplitude does not correspond
to the optical theorem and does not describes the reflection and
transmission coefficient. Namely
$|\clT|^2+|\clR|^2= 1+\frac{2\lambda\hbar^2 k}{(\lambda-\hbar^2k)^2}$
and it violets the law of conservation of the number of particles.
Therefore, the imaginary optical potential $i\delta(y)$
relates to the scattering with violation of the optical theorem.


\section*{References}

\begin{thebibliography}{99}

%

\bibitem{MaMa80} G. Matheron, G. de Marsily,
Is transport in porous media always diffusive? a
counterexample, Water Resour. Res. 16: 901 - 917, (1980).

\bibitem{BaIo04} E. Baskin, A. Iomin,
Superdiffusion on a comb structure,
Phys. Rev. Lett. 93: 120603, 2004.


\bibitem{BeKe99a} M.V. Berry, J.P. Keating,
$H = xp$ and the Riemann zeros, in: J.P. Keating, D.E.
Khmelnitskii, I.V. Lerner (Eds.), \textit{Supersymmetry and Trace Formulae:
Chaos and Disorder}, Kluwer, New York, 1999.

\bibitem{BeKe99b} M.V. Berry, J.P. Keating,
The Riemann zeros and eigenvalue asymptotics,
SIAM Rev. 41(2): 236 - 266, 1999.

\bibitem{Connes} A. Connes, Trace formula in noncommutative
geometry and the zeros of the Riemann zeta function,
Selecta Math. New Ser. 5: 29, 1999; math.NT/9811068.

\bibitem{Sierra} G. Sierra, $H = xp$ with interaction and the
Riemann zeros, Nucl. Phys. B 776 [PM]: 327 - 364, 2007.

\bibitem{AlFuFu76} V. de Alfaro, S. Fubini, and G. Furlan,
Conformal invariance in quantum mechanics,
Nuovo Cimento A 34: 569 - 612, 1976.

\bibitem{Jackiw} R. Jackiw,  in \textit{M. A. B. Beg Memorial Volume}
ed A. Ali and P. Hoodbhoy, World Scientific, Singapore, 1991

\bibitem{vanW} C. van Winter,
An increasing entropy for a free quantum particle,
J. Math. Phys. 39: 3600 - 3618, 1998.

\bibitem{Armitage} J. V. Armitage, The Riemann hypothesis
and the Hamiltonian of a quantum mechanical system,
in \textit{Number Theory and Dynamical Systems}, edited
by M. M. Dodson and J. A. G. Vickers,
Cambridge University Press, Cambridge, UK, 1989.

\bibitem{TwMi06} J. Twamley, G. J. Milburn,
The quantum Mellin transform,
New J. Phys. 8: 328, (2006).

\bibitem{BhKhLa95} R. K. Bhaduri, A. Khare, and J. Law,
Phase of the Riemann zeta function and the inverted harmonic
oscillator,
Phys. Rev. E 52: 486 - 491, 1995.

\bibitem{NoVo97} S. Nonnenmacher and A. Voros,
Eigenstate structures around a hyperbolic point,
J. Phys. A 30: 295 - 315, 1997.

\bibitem{BeVi03} G. P. Berman and M. Vishik,
Long time evolution of quantum averages near stationary points,
Phys. Lett. A 319: 352 - 359, 2003.

\bibitem{Io2013} A. Iomin,
Exponential spreading and singular behavior of
quantum dynamics near hyperbolic points,
Phys. Rev. E 87: 054901, 2013.

\bibitem{Io2009} A. Iomin,
Fractional-time quantum dynamics,
Phys. Rev. E 80: 022103, 2009.

\bibitem{SaIoKo2020} T. Sandev, A. Iomin, and L. Kocarev,
Hitting times in turbulent diffusion due to multiplicative noise,
upublished 2020.

\bibitem{PeZe98}  A. M. Perelomov and Y. B. Zel'dovich,
\textit{Quantum Mechanics - Selected Topics},
WS, Singapore, 1998.

\bibitem{KrLuMa14} P. L. Krapivsky, J. M. Luck, and
K. Mallick, Survival of classical and quantum particles
in the presence of traps,
J. Stat. Phys. 154: 1430 - 1460, 2014.

\bibitem{DeKhRo2003} R. N. Deb, A. Khare, and B. D. Roy,
Complex optical potentials and
pseudo-Hermitian Hamiltonians,
Phys. Lett. A 307: 215 - 221, 2003.

\bibitem{fok} V. A. Fock, \textit{Foundations of Quantum Mechanics},
Nauka, Moscow, 1976 (in Russian).

\bibitem{LL} L. D. Landau, E. M. Lifshitz, Quantum Mechanics,
Pergamon, New York, 1977.

\bibitem{BaEr55} H. Bateman and A. Erd\'elyi,
\textit{Higher Transcendental Functions}, [V. 1 -- 3],
McGraw-Hill, New York, 1953--1955.

\bibitem{MaHa08} A. M. Mathai and H. J. Haubold,
\textit{Special Functions for Applied Scientists},
Springer, NY, 2008.

\bibitem{IMH2018} A. Iomin, V. Mendez, and W. Horsthemke,
\textit{Fractional Dynamics in Comb-like Structures}
WS, Singapore, 2018.

\bibitem{AgMuBl10} E. Agliary, O. M\"ulken, and A. Blumen,
Continuous-time quantum walks and trapping,
Int. J. of Bif. 20; 271 - 279, 2010.

\bibitem{Bl88} S. M. Blinder, Green’s function and propagator
for the one-dimensional $\delta$-function potential,
Phys. Rev. A 37: 973 - 976, 1988.

\bibitem{BaEr54} H. Bateman and A. Erd\'elyi,
\textit{Tables of integral transforms}, [V. 1],
McGraw-Hill, New York, 1954.

\bibitem{SaIoKo2019} T. Sandev, A. Iomin, and L. Kocarev,
Random search on comb, J. Phys. A: Math. Theor. 52: 465001, 2019.

\bibitem{IoSa_MDPI} A. Iomin and T. Sandev, Fractional diffusion
to a Cantor set in 2D, unpublished.

\bibitem{LSRJIK2020} E. K. Lenzi, T. Sandev, H. V. Ribeiro,
P. Jovanovski, A. Iomin, and L. Kocarev,
Anomalous diffusion and random search in xyz-comb: exact results,
J. Stat. Mech. 2020: 053203, 2020.


\bibitem{Eb65} J. H. Eberly, Quantum scattering theory in one dimension,
Am. J. Phys. 33: 771 - 773, 1965.

\end{thebibliography}
\end{document}